\documentclass[11pt,oneside,reqno,english,american]{amsart}
\usepackage[T1]{fontenc}
\usepackage[latin9]{inputenc}
\usepackage{xcolor}
\usepackage{verbatim}
\usepackage{amsbsy}
\usepackage{amstext}
\usepackage{amsthm}
\usepackage{amssymb}
\usepackage{undertilde}
\usepackage{graphicx}
\usepackage{xargs}[2008/03/08]

\makeatletter
\numberwithin{equation}{section}
\numberwithin{figure}{section}
\theoremstyle{plain}
\newtheorem{thm}{\protect\theoremname}[section]
\theoremstyle{remark}
\newtheorem{rem}[thm]{\protect\remarkname}
\theoremstyle{definition}
\newtheorem{example}[thm]{\protect\examplename}

\usepackage{mathtools}
\usepackage{amsthm}
\usepackage{pdfsync} 
\usepackage{hyperref}
\usepackage[all]{xy}

 \theoremstyle{plain}
  
\usepackage[nf]{coelacanth}
\usepackage[stix2]{newtxmath}
\usepackage[cal=boondoxo,bb=boondox,frak=boondox]{mathalfa}
\usepackage{xcolor} 
\definecolor{brown(traditional)}{rgb}{0.59, 0.29, 0.0}
\definecolor{blue(ryb)}{rgb}{0.01, 0.28, 1.0}
\definecolor{red}{rgb}{1.0, 0.0, 0.0}
\definecolor{magenta}{rgb}{1.0, 0.0, 1.0}
\definecolor{mahogany}{rgb}{0.75, 0.25, 0.0}
\definecolor{lavenderpurple}{rgb}{0.59, 0.48, 0.71}
\definecolor{olive}{rgb}{0.5, 0.5, 0.0}
\definecolor{brickred}{rgb}{0.8, 0.25, 0.33}
\definecolor{antiquefuchsia}{rgb}{0.57, 0.36, 0.51}
\definecolor{bole}{rgb}{0.47, 0.27, 0.23}
\definecolor{darkolivegreen}{rgb}{0.33, 0.42, 0.18}
\definecolor{deepjunglegreen}{rgb}{0.0, 0.29, 0.29}
\definecolor{brickred}{rgb}{0.8, 0.25, 0.33}
\definecolor{deepjunglegreen}{rgb}{0.0, 0.29, 0.29}
\definecolor{darkpastelgreen}{rgb}{0.01, 0.75, 0.24}
\definecolor{green(pigment)}{rgb}{0.0, 0.65, 0.31}
\definecolor{junglegreen}{rgb}{0.16, 0.67, 0.53}
\definecolor{officegreen}{rgb}{0.0, 0.5, 0.0}
\definecolor{seagreen}{rgb}{0.18, 0.55, 0.34}
\definecolor{teal}{rgb}{0.0, 0.5, 0.5}
\definecolor{brightgreen}{rgb}{0.4, 1.0, 0.0}
\definecolor{electricgreen}{rgb}{0.0, 1.0, 0.0}
\definecolor{malachite}{rgb}{0.04, 0.85, 0.32}

\newcommand{\myline}{\smallskip{}
\begin{center}
\rule{0.6\textwidth}{0.3pt}
\end{center}
\par
\smallskip{}
}
\usepackage{undertilde}
\usepackage{accents}

\makeatother

\usepackage{babel}
\addto\captionsamerican{\renewcommand{\examplename}{Example}}
\addto\captionsamerican{\renewcommand{\remarkname}{Remark}}
\addto\captionsamerican{\renewcommand{\theoremname}{Theorem}}
\addto\captionsenglish{\renewcommand{\examplename}{Example}}
\addto\captionsenglish{\renewcommand{\remarkname}{Remark}}
\addto\captionsenglish{\renewcommand{\theoremname}{Theorem}}
\providecommand{\examplename}{Example}
\providecommand{\remarkname}{Remark}
\providecommand{\theoremname}{Theorem}

\begin{document}

\global\long\def\ga{\alpha}%
\global\long\def\gb{\beta}%
\global\long\def\ggm{\gamma}%
\global\long\def\go{\omega}%
\global\long\def\gs{\sigma}%
\global\long\def\gd{\delta}%
\global\long\def\gD{\Delta}%
\global\long\def\vph{\phi}%
\global\long\def\gf{\varphi}%
\global\long\def\gk{\kappa}%
\global\long\def\gl{\lambda}%
\global\long\def\gz{\zeta}%
\global\long\def\gh{\eta}%
\global\long\def\gy{\upsilon}%
\global\long\def\gth{\theta}%
\global\long\def\gO{\Omega}%
\global\long\def\gG{\Gamma}%

\global\long\def\eps{\varepsilon}%
\global\long\def\epss#1#2{\varepsilon_{#2}^{#1}}%
\global\long\def\ep#1{\eps_{#1}}%

\global\long\def\wh#1{\widehat{#1}}%
\global\long\def\hi{\hat{\imath}}%
\global\long\def\hj{\hat{\jmath}}%
\global\long\def\hk{\hat{k}}%
\global\long\def\ol#1{\overline{#1}}%
\global\long\def\ul#1{\underline{#1}}%

\global\long\def\spec#1{\textsf{#1}}%

\global\long\def\ui{\wh{\boldsymbol{\imath}}}%
\global\long\def\uj{\wh{\boldsymbol{\jmath}}}%
\global\long\def\uk{\widehat{\boldsymbol{k}}}%

\global\long\def\uI{\widehat{\mathbf{I}}}%
\global\long\def\uJ{\widehat{\mathbf{J}}}%
\global\long\def\uK{\widehat{\mathbf{K}}}%

\global\long\def\bs#1{\boldsymbol{#1}}%
\global\long\def\vect#1{\mathbf{#1}}%
\global\long\def\bi#1{\textbf{\emph{#1}}}%

\global\long\def\uv#1{\widehat{\boldsymbol{#1}}}%
\global\long\def\cross{\times}%

\global\long\def\ddt{\frac{\dee}{\dee t}}%
\global\long\def\dbyd#1{\frac{\dee}{\dee#1}}%
\global\long\def\dby#1#2{\frac{\partial#1}{\partial#2}}%
\global\long\def\dxdt#1{\frac{\dee#1}{\dee t}}%

\global\long\def\vct#1{\bs{#1}}%

\global\long\def\partialby#1#2{\frac{\partial#1}{\partial x^{#2}}}%
\newcommandx\parder[2][usedefault, addprefix=\global, 1=]{\frac{\partial#2}{\partial#1}}%

\global\long\def\fall{,\quad\text{for all}\quad}%

\global\long\def\reals{\mathbb{R}}%

\global\long\def\rthree{\reals^{3}}%
\global\long\def\rsix{\reals^{6}}%
\global\long\def\rn{\reals^{n}}%
\global\long\def\eucl{\mathbb{E}}%
\global\long\def\euthree{\eucl^{3}}%
\global\long\def\euln{\eucl^{n}}%

\global\long\def\prn{\reals^{n+}}%
\global\long\def\nrn{\reals^{n-}}%
\global\long\def\cprn{\overline{\reals}^{n+}}%
\global\long\def\cnrn{\overline{\reals}^{n-}}%
\global\long\def\rt#1{\reals^{#1}}%
\global\long\def\rtw{\reals^{12}}%

\global\long\def\les{\leqslant}%
\global\long\def\ges{\geqslant}%

\global\long\def\dee{\textrm{d}}%
\global\long\def\di{d}%
\global\long\def\dX{\dee\bp}%
\global\long\def\dx{\dee x}%
\global\long\def\D{D}%

\global\long\def\from{\colon}%
\global\long\def\tto{\longrightarrow}%
\global\long\def\lmt{\longmapsto}%
\global\long\def\lhr{\lhook\joinrel\longrightarrow}%
\global\long\def\mto{\mapsto}%

\global\long\def\abs#1{\left|#1\right|}%

\global\long\def\isom{\cong}%

\global\long\def\comp{\circ}%

\global\long\def\cl#1{\overline{#1}}%

\global\long\def\fun{\varphi}%

\global\long\def\interior{\textrm{Int}\,}%
\global\long\def\inter#1{\kern0pt  #1^{\mathrm{o}}}%
\global\long\def\interior{\textrm{Int}\,}%
\global\long\def\inter#1{\kern0pt  #1^{\mathrm{o}}}%
\global\long\def\into{\mathrm{o}}%

\global\long\def\sign{\textrm{sign}\,}%
\global\long\def\sgn#1{(-1)^{#1}}%
\global\long\def\sgnp#1{(-1)^{\abs{#1}}}%

\global\long\def\du#1{#1^{*}}%

\global\long\def\tsum{{\textstyle \sum}}%
\global\long\def\lsum{{\textstyle \sum}}%

\global\long\def\dimension{\textrm{dim}\,}%

\global\long\def\esssup{\textrm{ess}\,\sup}%

\global\long\def\ess{\textrm{{ess}}}%

\global\long\def\kernel{\mathop{\textrm{\textup{Kernel}}}}%

\global\long\def\support{\mathop{\textrm{\textup{supp}}}}%

\global\long\def\image{\mathop{\textrm{\textup{Image}}}}%

\global\long\def\diver{\mathop{\textrm{\textup{div}}}}%

\global\long\def\spanv{\textrm{span}}%

\global\long\def\tr{\mathop{\textrm{\textup{tr}}}}%
\global\long\def\tran{\mathrm{tr}}%

\global\long\def\opt{\mathrm{opt}}%

\global\long\def\resto#1{|_{#1}}%
\global\long\def\incl{\mathcal{I}}%
\global\long\def\iden{\imath}%
\global\long\def\idnt{\textrm{Id}}%
\global\long\def\rest{\rho}%
\global\long\def\extnd{e_{0}}%

\global\long\def\proj{\textrm{pr}}%

\global\long\def\L#1{L\bigl(#1\bigr)}%
\global\long\def\LS#1{L_{S}\bigl(#1\bigr)}%

\global\long\def\ino#1{\int_{#1}}%

\global\long\def\half{\frac{1}{2}}%
\global\long\def\shalf{{\scriptstyle \half}}%
\global\long\def\third{\frac{1}{3}}%

\global\long\def\empt{\varnothing}%

\global\long\def\paren#1{\left(#1\right)}%
\global\long\def\bigp#1{\bigl(#1\bigr)}%
\global\long\def\biggp#1{\biggl(#1\biggr)}%
\global\long\def\Bigp#1{\Bigl(#1\Bigr)}%

\global\long\def\braces#1{\left\{  #1\right\}  }%
\global\long\def\sqbr#1{\left[#1\right]}%
\global\long\def\anglep#1{\left\langle #1\right\rangle }%

\global\long\def\bigabs#1{\bigl|#1\bigr|}%
\global\long\def\dotp#1{#1^{\centerdot}}%
\global\long\def\pdot#1{#1^{\bs{\!\cdot}}}%

\global\long\def\eq{\sim}%
\global\long\def\quot{/\!\!\eq}%
\global\long\def\by{\!/\!}%

\global\long\def\stp{\text{\small\ensuremath{\bigodot}}}%
\global\long\def\tp{\text{\small\ensuremath{\bigotimes}}}%

\global\long\def\mi#1{#1}%
\global\long\def\mii{I}%
\global\long\def\mie#1#2{#1_{1}\cdots#1_{#2}}%

\global\long\def\smi#1{\boldsymbol{#1}}%
\global\long\def\asmi#1{#1}%
\global\long\def\ordr#1{\left\langle #1\right\rangle }%

\global\long\def\symm#1{\paren{#1}}%
\global\long\def\smtr{\mathcal{S}}%

\global\long\def\perm{p}%
\global\long\def\sperm{\mathcal{P}}%

\global\long\def\oneto{1,\dots,}%

\global\long\def\lisub#1#2#3{#1_{1}#2\dots#2#1_{#3}}%

\global\long\def\lisup#1#2#3{#1^{1}#2\dots#2#1^{#3}}%

\global\long\def\lisubb#1#2#3#4{#1_{#2}#3\dots#3#1_{#4}}%

\global\long\def\lisubbc#1#2#3#4{#1_{#2}#3\cdots#3#1_{#4}}%

\global\long\def\lisubbwout#1#2#3#4#5{#1_{#2}#3\dots#3\widehat{#1}_{#5}#3\dots#3#1_{#4}}%

\global\long\def\lisubc#1#2#3{#1_{1}#2\cdots#2#1_{#3}}%

\global\long\def\lisupc#1#2#3{#1^{1}#2\cdots#2#1^{#3}}%

\global\long\def\lisupp#1#2#3#4{#1^{#2}#3\dots#3#1^{#4}}%

\global\long\def\lisuppc#1#2#3#4{#1^{#2}#3\cdots#3#1^{#4}}%

\global\long\def\lisuppwout#1#2#3#4#5#6{#1^{#2}#3#4#3\wh{#1^{#6}}#3#4#3#1^{#5}}%

\global\long\def\lisubbwout#1#2#3#4#5#6{#1_{#2}#3#4#3\wh{#1}_{#6}#3#4#3#1_{#5}}%

\global\long\def\lisubwout#1#2#3#4{#1_{1}#2\dots#2\widehat{#1}_{#4}#2\dots#2#1_{#3}}%

\global\long\def\lisupwout#1#2#3#4{#1^{1}#2\dots#2\widehat{#1^{#4}}#2\dots#2#1^{#3}}%

\global\long\def\lisubwoutc#1#2#3#4{#1_{1}#2\cdots#2\widehat{#1}_{#4}#2\cdots#2#1_{#3}}%

\global\long\def\twp#1#2#3{\dee#1^{#2}\wedge\dee#1^{#3}}%

\global\long\def\thp#1#2#3#4{\dee#1^{#2}\wedge\dee#1^{#3}\wedge\dee#1^{#4}}%

\global\long\def\fop#1#2#3#4#5{\dee#1^{#2}\wedge\dee#1^{#3}\wedge\dee#1^{#4}\wedge\dee#1^{#5}}%

\global\long\def\idots#1{#1\dots#1}%
\global\long\def\icdots#1{#1\cdots#1}%

\global\long\def\norm#1{\|#1\|}%

\global\long\def\nonh{\heartsuit}%

\global\long\def\nhn#1{\norm{#1}^{\nonh}}%

\global\long\def\bigmid{\,\bigl|\,}%

\global\long\def\trps{^{{\scriptscriptstyle \textsf{T}}}}%

\global\long\def\testfuns{\mathcal{D}}%

\global\long\def\ntil#1{\tilde{#1}{}}%

\global\long\def\pis{y}%
\global\long\def\xo{\pis_{0}}%
\global\long\def\x{x}%

\global\long\def\pib{x}%
\global\long\def\bp{X}%
\global\long\def\ii{i}%
\global\long\def\ia{\alpha}%
\global\long\def\fp{y}%
\global\long\def\piv{v}%

\global\long\def\ib{i}%
\global\long\def\is{\alpha}%

\global\long\def\pbndo{\Gamma}%
\global\long\def\bndoo{\pbndo_{0}}%
 
\global\long\def\bndot{\pbndo_{t}}%
\global\long\def\intb{\inter{\body}}%
\global\long\def\bndb{\bdry\body}%

\global\long\def\cloo{\cl{\gO}}%

\global\long\def\nor{\mathbf{n}}%
\global\long\def\Nor{\mathbf{N}}%

\global\long\def\dA{\,\dee A}%

\global\long\def\dV{\,\dee V}%

\global\long\def\eps{\varepsilon}%

\global\long\def\tv{v}%
\global\long\def\av{u}%

\global\long\def\svs{\mathcal{W}}%
\global\long\def\vs{\mathbf{V}}%
\global\long\def\avs{\mathbf{U}}%
\global\long\def\affsp{\mathcal{A}}%
\global\long\def\man{\mathcal{M}}%
\global\long\def\odman{\mathcal{N}}%
\global\long\def\subman{\mathcal{V}}%
\global\long\def\pt{p}%

\global\long\def\vbase{e}%
\global\long\def\sbase{\mathbf{e}}%
\global\long\def\msbase{\mathfrak{e}}%
\global\long\def\vect{v}%
\global\long\def\dbase{\sbase}%

\global\long\def\chart{\varphi}%
\global\long\def\Chart{\Phi}%

\global\long\def\mind{\alpha}%
\global\long\def\vb{W}%
\global\long\def\vbp{\pi}%

\global\long\def\vbt{\mathcal{E}}%
\global\long\def\fib{\vs}%
\global\long\def\vbts{W}%
\global\long\def\avb{U}%
\global\long\def\vbp{\xi}%

\global\long\def\chart{\vph}%
\global\long\def\vbchart{\Phi}%

\global\long\def\jetb#1{J^{#1}}%
\global\long\def\jet#1{j^{1}(#1)}%
\global\long\def\tjet{\tilde{\jmath}}%

\global\long\def\Jet#1{J^{1}(#1)}%

\global\long\def\jetm#1{j_{#1}}%

\global\long\def\coj{\mathfrak{d}}%

\global\long\def\alt{\mathfrak{A}}%

\global\long\def\pou{\eta}%

\global\long\def\ext{{\textstyle \bigwedge}}%
\global\long\def\forms{\Omega}%

\global\long\def\dotwedge{\dot{\mbox{\ensuremath{\wedge}}}}%

\global\long\def\vel{\theta}%

\global\long\def\Jac{\mathcal{J}}%

\global\long\def\contr{\mathbin{\raisebox{0.4pt}{\mbox{\ensuremath{\lrcorner}}}}}%
\global\long\def\fcor{\llcorner}%
\global\long\def\bcor{\lrcorner}%
\global\long\def\fcontr{\mathbin{\raisebox{0.4pt}{\mbox{\ensuremath{\llcorner}}}}}%

\global\long\def\lie{\mathcal{L}}%

\global\long\def\ssym#1#2{\ext^{#1}T^{*}#2}%

\global\long\def\sh{^{\sharp}}%

\global\long\def\nfo{\ext^{n}T^{*}\base}%
\global\long\def\dfs{\ext^{d}T^{*}\base}%
\global\long\def\dmfs{\ext^{d-1}T^{*}\base}%

\global\long\def\spc{\mathcal{S}}%
\global\long\def\sptm{\mathcal{E}}%
\global\long\def\evnt{e}%
\global\long\def\frame{\Psi}%

\global\long\def\timeman{\mathcal{T}}%
\global\long\def\zman{t}%
\global\long\def\dims{n}%
\global\long\def\m{\dims-1}%
\global\long\def\dimw{m}%

\global\long\def\wc{z}%

\global\long\def\fourv#1{\mbox{\ensuremath{\mathfrak{#1}}}}%

\global\long\def\pbform#1{\undertilde{#1}}%
\global\long\def\util#1{\raisebox{-5pt}{\ensuremath{{\scriptscriptstyle \sim}}}\!\!\!#1}%

\global\long\def\utilJ{\util J}%

\global\long\def\utilRho{\util{\rho}}%

\global\long\def\body{\mathcal{B}}%
\global\long\def\man{\mathcal{M}}%
\global\long\def\var{\mathcal{V}}%
\global\long\def\base{\mathcal{X}}%
\global\long\def\fb{\mathcal{Y}}%
\global\long\def\srfc{\mathcal{Z}}%
\global\long\def\dimb{n}%
\global\long\def\dimf{m}%
\global\long\def\afb{\mathcal{Z}}%

\global\long\def\bdry{\partial}%

\global\long\def\gO{\varOmega}%

\global\long\def\reg{\mathcal{R}}%
\global\long\def\bdrr{\bdry\reg}%

\global\long\def\bdom{\bdry\gO}%

\global\long\def\bndo{\partial\gO}%

\global\long\def\tpr{\vartheta}%

\global\long\def\mot{M}%
\global\long\def\vf{w}%
\global\long\def\const{h}%

\global\long\def\avf{u}%

\global\long\def\stn{\varepsilon}%
\global\long\def\djet{\chi}%

\global\long\def\jvf{\eps}%

\global\long\def\rig{r}%

\global\long\def\rigs{\mathcal{R}}%

\global\long\def\qrigs{\!/\!\rigs}%

\global\long\def\qd{\!/\,\!\kernel\diffop}%

\global\long\def\dis{\chi}%
\global\long\def\conf{\kappa}%
\global\long\def\invc{\hat{\conf}^{-1}}%
\global\long\def\dinvc{\hat{\conf}^{-1*}}%
\global\long\def\csp{\mathcal{Q}}%

\global\long\def\embds{\textrm{Emb}}%

\global\long\def\lc{A}%

\global\long\def\lv{\dot{A}}%
\global\long\def\alv{\dot{B}}%

\global\long\def\j{\mathop{\mathrm{j}}}%
\global\long\def\mapp{M}%
\global\long\def\J{J}%
\global\long\def\jex{\mathop{}\!\mathrm{j}}%

\global\long\def\fc{F}%
\global\long\def\load{f}%
\global\long\def\afc{g}%

\global\long\def\bfc{\mathbf{b}}%
\global\long\def\bfcc{b}%

\global\long\def\sfc{\mathbf{t}}%
\global\long\def\sfcc{t}%

\global\long\def\stm{\varsigma}%
\global\long\def\std{S}%
\global\long\def\tst{\sigma}%
\global\long\def\tstd{s}%
\global\long\def\st{\sigma}%
\global\long\def\vst{\varsigma}%
\global\long\def\vstd{S}%
\global\long\def\tstm{\sigma}%
\global\long\def\vstm{\varsigma}%

\global\long\def\stp{S_{P}}%
\global\long\def\slf{R}%

\global\long\def\crel{\Phi}%

\global\long\def\stmat{\tau}%

\global\long\def\gdiv{\bdry\textrm{iv\,}}%
\global\long\def\extjet{\mathfrak{d}}%

\global\long\def\smc#1{\mathfrak{#1}}%

\global\long\def\nhs{P}%
\global\long\def\nhsa{P}%
\global\long\def\nhsb{\underline{P}}%

\global\long\def\soc{Z}%

\global\long\def\sts{\varSigma}%
\global\long\def\spstd{\mathfrak{S}}%
\global\long\def\sptst{\mathfrak{T}}%
\global\long\def\spnhs{\mathcal{P}}%
\global\long\def\Ljj{\L{J^{1}(J^{k-1}\vb),\ext^{n}T^{*}\base}}%

\global\long\def\spsb{\text{\Large\ensuremath{\Delta}}}%

\global\long\def\ened{\mathfrak{w}}%
\global\long\def\energy{\mathfrak{W}}%

\global\long\def\ebdfc{T}%
\global\long\def\optimum{\st^{\textrm{opt}}}%
\global\long\def\scf{K}%

\global\long\def\grp{G}%
\global\long\def\gact{A}%
\global\long\def\gid{e}%
\global\long\def\gel{\ggm}%

\global\long\def\ael{\upsilon}%
\global\long\def\lal{\mathfrak{g}}%

\global\long\def\prop{P}%
\global\long\def\expr{\Pi}%

\global\long\def\aprop{Q}%

\global\long\def\flux{\omega}%
\global\long\def\aflux{\psi}%

\global\long\def\fform{\tau}%

\global\long\def\dimn{n}%

\global\long\def\sdim{{\dimn-1}}%

\global\long\def\fdens{\phi}%

\global\long\def\pform{\varsigma}%
\global\long\def\vform{\beta}%
\global\long\def\sform{\tau}%
\global\long\def\flow{J}%
\global\long\def\n{\m}%
\global\long\def\cmap{\mathfrak{t}}%
\global\long\def\vcmap{\varSigma}%

\global\long\def\mvec{\mathfrak{v}}%
\global\long\def\mveco#1{\mathfrak{#1}}%
\global\long\def\mv#1{\mathfrak{#1}}%
\global\long\def\smbase{\mathfrak{e}}%
\global\long\def\spx{\simp}%
\global\long\def\il{l}%
\global\long\def\awe{\frown}%

\global\long\def\hp{H}%
\global\long\def\ohp{h}%

\global\long\def\hps{G_{\dims-1}(T\spc)}%
\global\long\def\ohps{G_{\dims-1}^{\perp}(T\spc)}%

\global\long\def\hyper{\mathcal{S}}%

\global\long\def\hpsx{G_{\dims-1}(\tspc)}%
\global\long\def\ohpsx{G_{\dims-1}^{\perp}(\tspc)}%

\global\long\def\fbun{F}%

\global\long\def\flowm{\Phi}%

\global\long\def\tgb{T\spc}%
\global\long\def\ctgb{T^{*}\spc}%
\global\long\def\tspc{T_{\pis}\spc}%
\global\long\def\dspc{T_{\pis}^{*}\spc}%

\global\long\def\fflow{\fourv J}%
\global\long\def\fvform{\mathfrak{b}}%
\global\long\def\fsform{\mathfrak{t}}%
\global\long\def\fpform{\mathfrak{s}}%
\global\long\def\lfc{\mathfrak{F}}%

\global\long\def\maxw{\mathfrak{g}}%
\global\long\def\frdy{\mathfrak{f}}%
\global\long\def\ptnl{\psi}%
\global\long\def\tptn{\Psi}%
\global\long\def\vptn{\mathfrak{a}}%
\global\long\def\mtst{\tstd_{M}}%
\global\long\def\mvst{\vstd_{M}}%

\global\long\def\sobp#1#2{W_{#2}^{#1}}%

\global\long\def\inner#1#2{\left\langle #1,#2\right\rangle }%

\global\long\def\fields{\sobp pk(\vb)}%

\global\long\def\bodyfields{\sobp p{k_{\partial}}(\vb)}%

\global\long\def\forces{\sobp pk(\vb)^{*}}%

\global\long\def\bfields{\sobp p{k_{\partial}}(\vb\resto{\bndo})}%

\global\long\def\loadp{(\sfc,\bfc)}%

\global\long\def\strains{\lp p(\jetb k(\vb))}%

\global\long\def\stresses{\lp{p'}(\jetb k(\vb)^{*})}%

\global\long\def\diffop{D}%

\global\long\def\strainm{E}%

\global\long\def\incomps{\vbts_{\yieldf}}%

\global\long\def\devs{L^{p'}(\eta_{1}^{*})}%

\global\long\def\incompsns{L^{p}(\eta_{1})}%

\global\long\def\testf{\mathcal{D}}%
\global\long\def\dists{\mathcal{D}'}%

\global\long\def\codiv{\boldsymbol{\partial}}%

\global\long\def\currof#1{\tilde{#1}}%

\global\long\def\chn{c}%
\global\long\def\chnsp{\mathbf{C}}%

\global\long\def\current{T}%
\global\long\def\curr{R}%

\global\long\def\curd{S}%
\global\long\def\curwd#1{\wh{#1}}%
\global\long\def\curnd#1{\wh{#1}}%

\global\long\def\contrf{{\scriptstyle \smallfrown}}%

\global\long\def\prodf{{\scriptstyle \smallsmile}}%

\global\long\def\form{\omega}%

\global\long\def\dens{\rho}%

\global\long\def\simp{s}%
\global\long\def\ssimp{\Delta}%
\global\long\def\cpx{K}%

\global\long\def\cell{C}%

\global\long\def\chain{B}%
\global\long\def\A{A}%
\global\long\def\B{B}%

\global\long\def\ach{A}%

\global\long\def\coch{X}%

\global\long\def\scale{s}%

\global\long\def\fnorm#1{\norm{#1}^{\flat}}%

\global\long\def\chains{\mathcal{A}}%

\global\long\def\ivs{\boldsymbol{U}}%

\global\long\def\mvs{\boldsymbol{V}}%

\global\long\def\cvs{\boldsymbol{W}}%

\global\long\def\ndual#1{#1'}%

\global\long\def\nd{'}%

\global\long\def\cee#1{C^{#1}}%

\global\long\def\lone{\{L^{1}\}}%

\global\long\def\linf{L^{\infty}}%

\global\long\def\lp#1{L^{#1}}%

\global\long\def\ofbdo{(\bndo)}%

\global\long\def\ofclo{(\cloo)}%

\global\long\def\vono{(\gO,\rthree)}%

\global\long\def\lomu{\{L^{1,\mu}\}}%
\global\long\def\limu{L^{\infty,\mu}}%
\global\long\def\limub{\limu(\body,\rthree)}%
\global\long\def\lomub{\lomu(\body,\rthree)}%

\global\long\def\vonbdo{(\bndo,\rthree)}%
\global\long\def\vonbdoo{(\bndoo,\rthree)}%
\global\long\def\vonbdot{(\bndot,\rthree)}%

\global\long\def\vonclo{(\cl{\gO},\rthree)}%

\global\long\def\strono{(\gO,\reals^{6})}%

\global\long\def\sob{\{W_{1}^{1}\}}%

\global\long\def\sobb{\sob(\gO,\rthree)}%

\global\long\def\lob{\lone(\gO,\rthree)}%

\global\long\def\lib{\linf(\gO,\reals^{12})}%

\global\long\def\ofO{(\gO)}%

\global\long\def\oneo{{1,\gO}}%
\global\long\def\onebdo{{1,\bndo}}%
\global\long\def\info{{\infty,\gO}}%

\global\long\def\infclo{{\infty,\cloo}}%

\global\long\def\infbdo{{\infty,\bndo}}%
\global\long\def\lobdry{\lone(\bdry\gO,\rthree)}%

\global\long\def\ld{LD}%

\global\long\def\ldo{\ld\ofO}%
\global\long\def\ldoo{\ldo_{0}}%

\global\long\def\trace{\gamma}%
\global\long\def\dtrace{\delta}%
\global\long\def\gtrace{\beta}%

\global\long\def\pr{\proj_{\rigs}}%

\global\long\def\pq{\proj}%

\global\long\def\qr{\,/\,\reals}%

\global\long\def\aro{S_{1}}%
\global\long\def\art{S_{2}}%

\global\long\def\mo{m_{1}}%
\global\long\def\mt{m_{2}}%

\global\long\def\ebdfc{T}%

\global\long\def\mini{\Omega}%
\global\long\def\optimum{s^{\mathrm{opt}}}%
\global\long\def\scf{K}%
\global\long\def\opsf{\st^{\mathrm{opt}}}%
\global\long\def\doptimum{s^{\opt,{\scriptscriptstyle D}}}%
\global\long\def\loptimum{s^{\opt,{\scriptscriptstyle \mathcal{M}}}}%

\global\long\def\fsubs{M}%

\global\long\def\yieldc{B}%

\global\long\def\yieldf{Y}%

\global\long\def\trpr{\pi_{P}}%

\global\long\def\devpr{\pi_{\devsp}}%

\global\long\def\prsp{P}%

\global\long\def\devsp{D}%

\global\long\def\ynorm#1{\|#1\|_{\yieldf}}%

\global\long\def\colls{\Psi}%

\global\long\def\aro{S_{1}}%
\global\long\def\art{S_{2}}%

\global\long\def\mo{m_{1}}%
\global\long\def\mt{m_{2}}%

\global\long\def\trps{^{\mathsf{T}}}%

\global\long\def\hb{^{\mathrm{hb}}}%

\global\long\def\yieldst{s_{Y}}%

\global\long\def\yieldc{B}%

\global\long\def\lcap{C}%

\global\long\def\yieldf{Y}%

\global\long\def\sphpr{\pi_{P}}%

\global\long\def\devpr{\pi_{\devsp}}%

\global\long\def\prsp{P}%

\global\long\def\devsp{D}%

\global\long\def\ynorm#1{\|#1\|_{\yieldf}}%

\global\long\def\colls{\Psi}%

\global\long\def\cone{Q}%
\global\long\def\fpr{\Pi}%
\global\long\def\fprd{\fpr_{\devsp}}%
\global\long\def\fprp{\fpr_{\prsp}}%
\global\long\def\find{I_{\devsp}}%
\global\long\def\finp{I_{\prsp}}%

\global\long\def\rig{r}%
\global\long\def\rigs{\mathcal{R}}%
\global\long\def\qrigs{\!/\!\rigs}%
\global\long\def\anv{\omega}%
\global\long\def\I{I}%
\global\long\def\mone{M_{1}}%

\global\long\def\bd{BD}%

\global\long\def\po{\proj_{0}}%
\global\long\def\normp#1{\norm{#1}'_{\ld}}%

\global\long\def\ssx{S}%

\global\long\def\smap{s}%

\global\long\def\smat{\chi}%

\global\long\def\sx{e}%

\global\long\def\snode{P}%
\global\long\def\newmacroname{}%

\global\long\def\elem{e}%

\global\long\def\nel{L}%

\global\long\def\el{l}%

\global\long\def\gr{g}%
\global\long\def\ngr{G}%

\global\long\def\eldof{\alpha}%

\global\long\def\glbs{\psi}%

\global\long\def\ipln{\phi}%

\global\long\def\ndof{D}%

\global\long\def\dof{d}%

\global\long\def\nldof{N}%

\global\long\def\ldof{n}%

\global\long\def\lvf{\chi}%

\global\long\def\amat{A}%
\global\long\def\bmat{B}%

\global\long\def\subsp{\mathcal{M}}%
\global\long\def\zerofn{Z}%

\global\long\def\snomat{E}%

\global\long\def\femat{E}%

\global\long\def\tmat{T}%

\global\long\def\fvec{f}%

\global\long\def\snsp{\mathcal{S}}%

\global\long\def\slnsp{\Phi}%
\global\long\def\dslnsp{\Phi^{{\scriptscriptstyle D}}}%

\global\long\def\ro{r_{1}}%

\global\long\def\rtwo{r_{2}}%

\global\long\def\rth{r_{3}}%

\global\long\def\fmax{M}%

\global\long\def\dform{\psi}%

\global\long\def\srfc{\mathcal{S}}%

\global\long\def\semib{\mathrm{SB}}%

\global\long\def\tm#1{\overrightarrow{#1}}%
\global\long\def\tmm#1{\underrightarrow{\overrightarrow{#1}}}%

\global\long\def\itm#1{\overleftarrow{#1}}%
\global\long\def\itmm#1{\underleftarrow{\overleftarrow{#1}}}%

\global\long\def\ptrac{\mathcal{P}}%

\global\long\def\nh#1{\hat{#1}}%
\global\long\def\nj{\hat{\jmath}}%
\global\long\def\nJ{\hat{J}}%
\global\long\def\rin#1{\mathfrak{#1}}%
\global\long\def\npi{\hat{\pi}}%
\global\long\def\rp{\rin p}%
\global\long\def\rq{\rin q}%
\global\long\def\rr{\rin r}%

\global\long\def\xty{(\base,\fb)}%
\global\long\def\xts{(\base,\spc)}%
\global\long\def\r{r}%
\global\long\def\ntm{(\reals^{n},\reals^{m})}%

\global\long\def\tproj{\frame_{\timeman}}%
\global\long\def\sproj{\frame_{\spc}}%

\global\long\def\mtn{e}%
\global\long\def\sppp{\lambda}%

\global\long\def\mtsp{\mathscr{E}}%

\global\long\def\disp{g}%
\global\long\def\diffs{G}%

\global\long\def\bv{BV}%

\title[Growth to Fractals]{On the Evolution During Growth of Regular Boundaries of Bodies into
Fractals}
\author{Vladimir Goldshtein$\vphantom{N^{2}}^{1}$  and Reuven Segev$\vphantom{N^{2}}^{2}$}
\address{}
\keywords{Continuum mechanics; growing bodies; fractals; surface growth; de
Rham currents; flat chains.}
\begin{abstract}
Generalizing smooth volumetric growth to the singular case, using
de Rham currents and flat chains, we demonstrate how regular boundaries
of bodies may evolve to fractals. 
\end{abstract}

\date{\today\\[2mm]
$^1$ Department of Mathematics, Ben-Gurion University of the Negev, Israel. Email: vladimir@bgu.ac.il\\
$^2$ Department of Mechanical Engineering, Ben-Gurion University of the Negev, Israel. Email: rsegev@post.bgu.ac.il}
\subjclass[2000]{70A05; 74A05.}

\maketitle

\section{Introduction}

Fractals are limiting geometric objects obtained by rescaling an initial
object for scales that tend to zero. A fractal can be constructed
by geometric iterations as a complex invariant object from the solution
of a dynamical system (so-called a strange attractor) that is usually
studied by probabilistic methods.

Fractal boundaries of domains in $\rn$ can have a dimension that
is larger than $n-1$. More precisely, the Hausdorff dimension of
a fractal boundary can be any number between $n$ and $n-1$. The
rate of growth of the measure of the fractal boundary during its construction
by iterations or its evolution in time, depends directly on its Hausdorff
dimension. For example, the Hausdorff dimension of the boundary of
the classical von Koch snowflake is $\ln4/\ln3$.

Fractals often serve as mathematical models for complex geometries.
While models of physical and biological systems may not apply at arbitrarily
small scales, modeling such systems by fractals provides a convenient
framework and obviates the need to specify the range of scales where
the geometric model applies.

\begin{figure}
\begin{centering}
\includegraphics{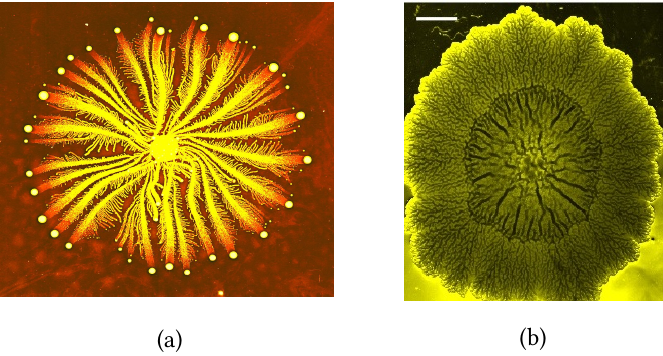}
\par\end{centering}
\caption{\label{fig:bact} Fractal-like bacteria colonies: \protect \\
(a) \emph{Paenibacillus vortex} sp. bacteria. (By Eshel Ben-Jacob,
\texttt{https://commons.wikimedia.org/wiki/File:}~\protect \\
\texttt{Paenibacillus\_vortex\_colony.jpg.)} \protect \\
(b) Surface morphologies of 14 days old WT Bacillus \emph{subtilis
}biofilm. Taken from \cite{Rafi2022} with permission of the authors.}
\end{figure}

Growth processes sometimes exhibit fractal-like behavior. The growth
patterns during phase transition (see, e.g., \cite{Suzuki-83,Stinchcombe89}),
growth patterns in trees, \cite{FractalsEverywhere}, and growth of
bacterial colonies, e.g., \cite{Fujikawa89,Obert90,Fujikawa91} (see
Figure \ref{fig:bact}), provide examples of such processes. Fractal-like
boundaries imply larger a surface area relative to the volume for
the transport of nutrients.

The mechanics of growth from the point of view of continuum mechanics
has gained attention since the end of the 20th century (see \cite{Skalak1982,Taber1995,Sev-Epst-GrowingBodies96}).
Growth and the possible creation and destruction of body points seem
to contradict the principle of material impenetrability of continuum
mechanics.

Surface growth, an important mode of growth, and its consequences
in terms of stresses and forces (for example, loc. cit, \cite{SozioYavari2016,Trusk-Zurlo-19,Pradhan-Yavari23}),
offers a particular challenge in the study of continuum theories of
growth. While theories of surface growth usually consider the evolution
of smooth surfaces, this paper considers the evolution of fractal-like
surfaces, specifically, the evolution of polyhedral and smooth surfaces
to fractals.

In this paper, some fractals are modeled mathematically as de Rham
currents. This is a natural extension of smooth volumetric growth
to the singular case. In the general formulation, this setting applies
in the general framework of proto-Galilean spacetime, a general fiber
bundle over the time axis (see \cite{Segev2022,goldshtein2023notes}).
As a typical example, we consider the von Koch snowflake, modeled
by a flat chain as in Whitney's geometric integration theory, \cite{Whitney1957}.

As background, we review in Section \ref{sec:Smooth-Growth} the basic
notions of smooth flux theory and volumetric growth in a proto-Galilean
spacetime, as in \cite{goldshtein2023notes}. Special attention is
given to the role played by a frame, a trivialization, as opposed
to frame invariant variables and operations. The theory is extended
to the singular case in Section \ref{sec:Currents} using the theory
of de Rham currents. We briefly review the basic notions of de Rham
currents, propose the relevant generalization of smooth volumetric
growth to the singular case, and give the example of surface growth
as singular volumetric growth.

Whitney's geometric integration theory for flat chains \cite{Whitney1957}
may be presented as a special case of the theory of currents, as in
\cite{Federer1969}. We present the fundamental ideas in Section \ref{sec:Whitey}.
Flat chains may be used to model various fractals, and we outline
the construction of the von Koch snowflake as an example.

The continuous evolution of a polyhedron to a fractal, the snowflake,
is demonstrated in Section \ref{sec:Fractal-Growth}.

Finally, in Section \ref{sec:Evolution-to-Fractals}, we use the theory
of conformal mappings and prime ends to propose a construction of
a smooth evolution of a two-dimensional region having a smooth boundary
into a fractal.

\section{\label{sec:Smooth-Growth}Smooth Growth in a Proto-Galilean Spacetime}

This section describes the geometric setting of spacetime for what
follows and presents the basic objects used to describe mathematically
smooth volumetric growth. In our formulation, smooth volumetric growth
arise from a nonvanishing source of an extensive property, mass, for
example. Body points corresponding to the extensive property under
consideration, may be defined. Since one considers time-evolution
of the extensive property, a specific model of spacetime should be
used.

\subsection{Proto-Galilean spacetime}

We use a generalized model of Galilean spacetime, proto-Galilean spacetime
(see \cite{Segev2022}). It is classical in the sense that to each
event $\evnt$ there corresponds a unique time $t=\pi(e)$. Thus,
if $\sptm$ denotes spacetime, and taking the time axis as $\reals$
for the sake of simplicity, there is a mapping 
\begin{equation}
\pi:\sptm\tto\reals.
\end{equation}
For each time $t\in\reals$, the inverse image $\sptm_{t}:=\pi^{-1}\{t\}$,
the collection of simultaneous events at $t$, is assumed to be diffeomorphic
with an $n$-dimensional space oriented manifold $\spc$. However,
in accordance with the Galilean point of view that location in space
is not absolute, there is no unique oriented preserving diffeomorphism
\begin{equation}
\Phi_{t}:\sptm_{t}\tto\spc.
\end{equation}

Spacetime $\sptm$ is assumed to be an $(n+1)$-dimensional manifold
having a fiber bundle structure provided by a collection of (local)
frames $\{(U_{a},\Phi_{a})\}$, such that the collection $\{U_{a}\}$
is an open cover of $\reals$ and 
\begin{equation}
\Phi_{a}:\pi^{-1}(U_{\ga})\tto U_{a}\times\spc,\qquad\text{such that}\qquad\proj_{1}\comp\Phi_{a}=\pi\label{eq:frame}
\end{equation}
are diffeomorphisms. Here, $\proj_{1}$ and $\proj_{2}$ denote the
projections on the first and second factors of the Cartesian product,
respectively. Thus, simply, a frame associates a particular point
$x\in\spc$ with each event $\evnt$, and 
\begin{equation}
\Phi_{a}:\evnt\lmt(t,x).
\end{equation}
Given, another frame,
\begin{equation}
\Phi_{b}:\pi^{-1}(U_{b})\tto U_{b}\times\spc,\qquad\evnt\lmt(t'=t,x')\label{eq:another-frame}
\end{equation}
such that $U_{a}\cap V_{b}\ne\varnothing$, $t\in U_{a}\cap V_{b}$,
there is a transformation
\begin{equation}
\Phi_{ba}(t):=\Phi_{b}\comp\Phi_{a}^{-1}\resto t:\spc\tto\spc,\qquad x\lmt x'.
\end{equation}

If $x^{i}$ are coordinates in a neighborhood of $x$, and $x'^{j}$
are coordinates in a neighborhood of $x'$, the mapping $\Phi_{ba}$
is represented by functions
\begin{equation}
x'^{j}=x'^{j}(t,x^{i}).\label{eq:transformation-components}
\end{equation}

The tangent mapping of the time-projection is
\begin{equation}
T\pi:T\sptm\tto T\reals.
\end{equation}
Let $\bs 1$be the natural tangent base vector to the time axis and
let $\bs 1^{*}$ be the dual element of $T^{*}\reals$. Then, a natural
one form $\dee t$ is induced in spacetime by the pullback of forms
as
\begin{equation}
\dee t=T^{*}\pi(\bs 1^{*}).
\end{equation}
Evidently, $\dee t$ is frame-independent.

Another frame-independent notion is that of a spacelike tangent vector.
A vector $v\in T\sptm$ is said to be spacelike if it is vertical,
that is, if
\begin{equation}
T\pi(v)=0.
\end{equation}
Note that if $v$ is spacelike, 
\begin{equation}
\begin{split}\dee t(v) & =T^{*}\pi(\bs 1^{*})(v),\\
 & =\bs 1^{*}(T\pi(v)),\\
 & =0.
\end{split}
\end{equation}
Conversely, if $\dee t(v)=0$, $v$ is spacelike. The collection of
spacelike tangent vectors, an $n$-dimensional subbundle of $T\sptm$
will be denoted as $V\sptm$. With some abuse of notation, we write
\begin{equation}
\tau:V\sptm\tto\sptm,
\end{equation}
for the restriction of 
\begin{equation}
\tau:T\sptm\tto\sptm,
\end{equation}
and we write
\begin{equation}
\incl_{V}:V\sptm\tto T\sptm\label{eq:incl_vertical}
\end{equation}
 for the inclusion mapping of the subbundle.

For any time $t\in\reals$, a spacelike vector is tangent to the $\sptm_{t}$
and the vector bundle $(V\sptm)_{t}$ is identical to $T(\sptm_{t})$
\begin{center}
\smallskip{}
\rule[0.5ex]{0.6\textwidth}{0.3pt}
\par\end{center}

\smallskip{}

We now consider additional objects available once a frame, $\Phi$,
on spacetime is given. To simplify the notation, we write the expressions
for the case where the frame is global in the sense that its domain
is $\sptm$. The tangent mapping 
\begin{equation}
T\Phi:T\sptm\tto T(\reals\times\spc)=T\reals\times T\spc
\end{equation}
is a vector bundle isomorphism. The components of $T\Phi$, are
\begin{equation}
T\Phi_{1}:T\sptm\tto T\reals,\quad v\lmt v^{t},\quad\text{and}\quad T\Phi_{2}:T\sptm\tto T\spc,\quad v\lmt v_{x}.
\end{equation}

Thus, we may define the vector field
\begin{equation}
\bdry_{t}=\parder[t]{}:=T\Phi^{-1}(\bs 1,0).
\end{equation}
It is emphasized that the ``timelike'' unit vector $\bdry_{t}$ is
frame-dependent as it requires one to keep the location in space ``fixed''
for distinct times. Note that
\begin{equation}
\begin{split}\dee t(\bdry_{t}) & =T^{*}\pi(\bs 1^{*})(T\Phi^{-1}(\bs 1,0)),\\
 & =\bs 1^{*}(T\pi\comp T\Phi^{-1}(\bs 1,0)),\\
 & =\bs 1^{*}(T(\pi\comp\Phi^{-1})(\bs 1,0)),\\
 & =\bs 1^{*}(T\proj_{1}(\bs 1,0)),\\
 & =\bs 1^{*}(\bs 1),
\end{split}
\end{equation}
where in the fourth line we used (\ref{eq:frame}), which implies
\begin{equation}
\begin{split}\pi\comp\Phi^{-1} & =\proj_{1}\comp\Phi\comp\Phi^{-1},\\
 & =\proj_{1}
\end{split}
\end{equation}
Thus, while $\bdry_{t}$ depends on the frame,
\begin{equation}
\dee t(\bdry_{t})=1
\end{equation}
independently of the frame. This also implies that $\bdry_{t}$ is
not spacelike, and that it complements the vertical subbundle to the
tangent bundle $T\sptm$.

In addition, it follows from (\ref{eq:frame}) that for any $v\in T\sptm$,
we have,
\begin{equation}
T\pi(v)=T(\proj_{1}\comp\Phi)(v)=T\proj_{1}\comp T\Phi(v),
\end{equation}
thus, $v$ is vertical if and only if 
\begin{equation}
v^{t}=T\Phi_{1}(v)=0.
\end{equation}
We conclude that every vector $v\in T\sptm$ is represented locally
as
\begin{equation}
v=v^{t}\bdry_{t}+v_{x}^{i}\bdry_{i},
\end{equation}
and every vertical vector is of the local form
\begin{equation}
v=v_{x}^{i}\bdry_{i}.
\end{equation}

Let $\Phi'$ be another frame so that the transformation rule is represented
locally as in (\ref{eq:transformation-components}),
\begin{equation}
(t',x'^{j})=(t'=t,x'^{j}(t,x^{i})),\qquad i,j=\oneto n.
\end{equation}
Hence, using the summation convention,
\begin{equation}
\bdry'_{t}=\parder[t']{}=\parder[t']t\parder[t]{}+\parder[t']{x^{i}}\parder[x^{i}]{},
\end{equation}
and writing $\bdry_{i}:=\bdry/\bdry x^{i}$ for the locally induced
base vectors, we conclude that 
\begin{equation}
\bdry'_{t}=\bdry_{t}+\parder[t']{x^{i}}\bdry_{i}.\label{eq:dt'}
\end{equation}

\subsection{Smooth extensive properties and flux fields}

We consider the fields associated with a smoothly distributed extensive
property $\expr$ in spacetime.

We start with the description of the source field, $\fourv s$, in
spacetime. Let $\fourv s$ be an $(n+1)$-form on spacetime. Given
a frame $\Phi$ in spacetime, $\fourv s$ is represented locally in
the form
\begin{equation}
\fourv s=\fourv s_{t1\dots n}\dee t\wedge\dee x,\qquad\text{where},\qquad\dee x:=\lisup{\dee x}{\wedge}n.
\end{equation}
As discussed above, the frame induces the unit timelike vector $\bdry_{t}$,
and so the contraction $\bdry_{t}\contr\fourv s$ is an $n$-form
is space. Using the inclusion mapping of the vertical subbundle as
in (\ref{eq:incl_vertical}), we may define the source $n$-form $\pform$
on $V\sptm$, as
\begin{equation}
\pform:=\incl_{V}^{*}(\bdry_{t}\contr\fourv J).
\end{equation}
Locally, $\pform$ is represented by
\begin{equation}
\pform=\fourv s_{t1\dots n}\dee x.
\end{equation}
The restriction of $\pform$ to $\forms^{n}((V\sptm)_{t}^{*})=\forms^{n}(T^{*}\sptm_{t})$,
denoted by $\pform(t)$ is a time-dependent $n$-form on $\sptm_{t}$.
It is interpreted as the source form in the instantaneous space, $\sptm_{t}$,
and for vertical vectors $\lisub v,n\in T(\sptm_{t})$, the evaluation
$\pform(t)(\lisub v,n)$ is interpreted as the infinitesimal amount
of the property produced in the infinitesimal element determined by
the vectors $v_{i}$ during a unit time interval, in agreement with
the standard interpretation of the source field.

If another frame $\Phi'$ is given, represented locally by the coordinates
$(t',x'^{j})$, then, using (\ref{eq:dt'}), the induced source on
spacelike vectors, $\pform'$ is represented locally by
\begin{equation}
\begin{split}\pform'=\incl^{*}(\bdry'_{t}\contr\fourv s) & =\left(\bdry_{t}+\parder[t']{x^{i}}\bdry_{i}\right)\contr\fourv s,\\
 & =\bdry_{t}\contr\fourv s+\parder[t']{x^{i}}\bdry_{i}\contr\fourv s,
\end{split}
\end{equation}
so that
\begin{equation}
\pform'=\pform+\parder[t']{x^{i}}\bdry_{i}\contr\fourv s.
\end{equation}

\myline

Next, we consider the flux field, $\fourv J$, on spacetime. Let $\fourv J$
be an $n$-form on spacetime. The restriction, 
\begin{equation}
\dens:=\incl_{V}^{*}(\fourv J),\label{eq:dens-from-J}
\end{equation}
of $\fourv J$ to vertical vectors is an $n$-form on $V\sptm$. For
an instant $t$, the restriction $\dens(t)$, of $\dens$ to $\forms^{n}((V\sptm)_{t}^{*})=\forms^{n}(T^{*}\sptm_{t})$
is a time dependent $n$-form on $\sptm_{t}$. It is interpreted as
the density of the property $\expr$ at the instantaneous space $\sptm_{t}$.
Note that while $\pform(t)$ is frame-dependent, $\dens(t)$ is frame-independent.

It follows that under a given frame, $\fourv J$ is represented locally
in the form
\begin{equation}
\begin{split}\fourv J & =\fourv J_{1\dots n}\dee x+\fourv J_{t1\dots\wh{\imath}\dots n}\dee t\wedge\dee x^{1}\wedge\cdots\wedge\wh{\dee x^{i}}\wedge\cdots\wedge\dee x^{n},\\
 & =\dens_{1\dots n}\dee x+\fourv J_{t1\dots\wh{\imath}\dots n}\dee t\wedge\dee x^{1}\wedge\cdots\wedge\wh{\dee x^{i}}\wedge\cdots\wedge\dee x^{n},
\end{split}
\label{eq:spacetime_flux}
\end{equation}
where $\dens_{1\dots n}=\fourv J_{1\dots n}$, and a hat indicates
the omission of a term.

We define the frame dependent
\begin{equation}
\flow:=-\bdry_{t}\contr\fourv J,\label{eq:J-space}
\end{equation}
so that locally,
\begin{equation}
\flow=\flow_{1\dots\wh{\imath}\dots n}\dee x^{1}\wedge\cdots\wedge\wh{\dee x^{i}}\wedge\cdots\wedge\dee x^{n},\qquad\flow_{1\dots\wh{\imath}\dots n}=-\fourv J_{t1\dots\wh{\imath}\dots n}.\label{eq:J-space-local}
\end{equation}
The restriction of $\flow$ to $\forms^{n-1}((V\sptm)_{t}^{*})=\forms^{n-1}(T^{*}\sptm_{t})$,
denoted by $\flow(t)$ is an $(n-1)$-form on $\sptm_{t}$. It is
interpreted as the flow, or flux, $(n-1)$-form. For vertical vectors
$\lisub v,{n-1}\in T(\sptm_{t})$, the evaluation $\pform(t)(\lisubb v1,{n-1})$
is interpreted as the amount of the property flowing out of the infinitesimal
hyperplane determined by the vectors $\lisubb v1,{n-1}$ for a unit
of time. This is in accordance with the traditional interpretation
of the flux field. We conclude that
\begin{equation}
\fourv J=\dens-\dee t\wedge\flow.\label{eq:spacetime_flux-vs_space_flux}
\end{equation}

For another frame, $\Phi'$, as above, another ``space''-flow $\flow'$
will be induced by $\bdry'_{t}$ as given by Equations (\ref{eq:dt'})
and (\ref{eq:J-space}). That is,
\begin{equation}
\begin{split}\flow' & =-\bdry'_{t}\contr\fourv J,\\
 & =-\left(\bdry_{t}+\parder[t']{x^{i}}\bdry_{i}\right)\contr\fourv J,\\
 & =\flow-\parder[t']{x^{i}}\bdry_{i}\contr\fourv J.
\end{split}
\end{equation}

\subsection{The balance equations in spacetime}

Evidently, the exterior differential $\dee\fourv J$ of the flux form
in spacetime has a frame-invariant meaning. For a given frame, differentiating
Equation (\ref{eq:spacetime_flux-vs_space_flux}),
\begin{equation}
\begin{split}\dee\fourv J & =\dee\dens-\dee(\dee t\wedge\flow),\\
 & =\dee\dens+\dee t\wedge\dee J.
\end{split}
\label{eq:dJ-frame}
\end{equation}
We may also differentiate Equation (\ref{eq:spacetime_flux}) to obtain
a specific local expression,
\begin{equation}
\begin{split}\dee\fourv J & =\frac{\bdry\dens_{1\dots n}}{\bdry t}\dee t\wedge\dee x+\parder[x^{j}]{\fourv J_{t1\dots\wh{\imath}\dots n}}\dee x^{j}\wedge\dee t\wedge\dee x^{1}\wedge\cdots\wedge\wh{\dee x^{i}}\wedge\cdots\wedge\dee x^{n},\\
 & =\dot{\dens}_{\oneto n}\dee t\wedge\dee x-\parder[x^{j}]{\fourv J_{t1\dots\wh{\imath}\dots n}}\dee t\wedge\dx^{j}\wedge\dee x^{1}\wedge\cdots\wedge\wh{\dee x^{i}}\wedge\cdots\wedge\dee x^{n},\\
 & =\dot{\dens}_{\oneto n}\dee t\wedge\dee x-\sum_{i=1}^{n}(-1)^{i-1}\parder[x^{i}]{\fourv J_{t1\dots\wh{\imath}\dots n}}\dee t\wedge\dee x,
\end{split}
\end{equation}
where a superimposed dot indicates partial time-differentiation. We
conclude that locally,
\begin{equation}
\dee\fourv J=\biggl(\dot{\dens}_{\oneto n}-\sum_{i=1}^{n}(-1)^{i-1}\parder[x^{i}]{\fourv J_{t1\dots\wh{\imath}\dots n}}\biggr)\dee t\wedge\dee x,\label{eq:dJ-components}
\end{equation}
in accordance with (\ref{eq:dJ-frame}).

As the balance for the property $\expr$ we put forward the frame-independent
equation
\begin{equation}
\dee\fourv J=\fourv s.\label{eq:Diff-Balance-smooth}
\end{equation}
Given a frame and local coordinates, the balance equation assumes
the form
\begin{equation}
\dot{\dens}_{\oneto n}-\sum_{i=1}^{n}(-1)^{i-1}\parder[x^{i}]{\fourv J_{t1\dots\wh{\imath}\dots n}}=\fourv s_{t1\dots n},
\end{equation}
or using (\ref{eq:J-space-local}),
\begin{equation}
\dot{\dens}_{\oneto n}+\sum_{i=1}^{n}(-1)^{i-1}\parder[x^{i}]{\flow_{1\dots\wh{\imath}\dots n}}=\fourv s_{t1\dots n},\label{eq:diff-balance-J}
\end{equation}

To clarify the relation between the foregoing equation and the standard
differential balance law of continuum mechanics, we first note that
if a volume element $\vel$ is given on $\sptm_{t}$, there is a unique
vertical vector field $\avf$ on $\sptm_{t}$ such that
\begin{equation}
\flow=\avf\contr\vel.
\end{equation}
Note that if $\dens(t)$ is positive everywhere, one may take it as
a natural volume element. If $\vel$ is represented locally as
\begin{equation}
\vel=\vel_{1\dots n}\dx,
\end{equation}
then the components of $\avf$ are given by
\begin{equation}
\avf^{i}=\sgn{i-1}\frac{\flow_{1\dots\wh i\dots n}}{\vel_{1\dots n}}.
\end{equation}
Thus, Equation (\ref{eq:diff-balance-J}) may be written in the form
\begin{equation}
\dot{\dens}_{\oneto n}+\parder[x^{i}]{(\vel_{\oneto n}\avf^{i})}=\fourv s_{t1\dots n}.
\end{equation}

\subsection{Body points}

Assume that $\fourv J$ does not vanish anywhere in $\sptm$. For
example, this will be satisfied if $\dens$ does not vanish, or alternatively,
if we restrict ourselves to an $(n+1)$-dimensional submanifold of
$\sptm$ where $\fourv J$ does not vanish.

At each event, $\evnt$, such that $\fourv J(\evnt)\ne0$, $\fourv J$
determines a one-dimensional subspace $D_{\evnt}\subset T_{\evnt}\sptm$.
A vector $v$ is in $D_{\evnt}$ if 
\begin{equation}
v\contr\fourv J=0,
\end{equation}
which means that the flux vanishes through any infinitesimal hyperplane
that contains $v$.

Alternatively, the vectors $v_{\evnt}\in D_{\eps}$ are those satisfying
\begin{equation}
\fourv J(\evnt)=v_{\evnt}\contr\vartheta(e)\label{eq:v_contr-theta}
\end{equation}
for some volume element $\vartheta$ on spacetime. (See \cite{Segev_Book_2023}
for further details.) As such, the form $\fourv J$ induces a one-dimensional
distribution, a subbundle, $D$, of the tangent bundle $T\sptm$.

Since any one-dimensional distribution is integrable, $\sptm$ is
the union of a disjoint family of one-dimensional submanifolds, a
folliation. Alternatively, each such one-dimensional submanifold can
be obtained as the integral line of the vector field $v$ satisfying
the condition $\fourv J=v\contr\vartheta$, or some volume element
$\vartheta$. It can be shown that as a submanifold, the integral
line is independent of the volume element chosen (see \cite{Segev_Book_2023}).

We view such a one-dimensional submanifold as a worldline, the trajectory
of a particle in spacetime. Thus, we identify a body point with such
an integral line. In other words, a body point is an equivalence class
of events using the equivalence relation $\sim$ defined as $\evnt\sim\evnt'$
if $\evnt$ and $\evnt'$ belong to the same integral submanifold.
The growing body is accordingly defined as $\sptm/\sim$.

\selectlanguage{english}%
The existence of body points is a proto-Newtonian idealization that
is convenient to classical continuum mechanics. For growing bodies
(for example, biological bodies), it can lead to contradictions. Every
cell in such bodies has a minimal and maximal size when it is alive.
This size can fluctuate during its evolution. Outside of these maximal
and minimal values, cells cannot exist. It is essential for so-called
fractal objects that are limiting objects in the proto-Newtonian sense
obtained by some infinite iterations or a similar procedure. If it
is a natural process, the main question is which iteration is final.
This is an open problem, and we will try below to discuss the evolution
in the fractal context.
\selectlanguage{american}%

\section{\label{sec:Currents}Modeling Singular Growth by Currents: Asymptotic
Version}

For the foregoing material, the smoothness of the flux $n$-form,
$\fourv J$, was essential. Smooth forms are generalized to possibly
singular objects by viewing them as continuous linear functionals
on locally convec topological vector spaces of infinitely smooth forms\textemdash de
Rham currents (see \cite{deRham1955}).

As a basic example, Let $\man$ be an $n$-dimensional manifold and
let $\gf$ be an $(n-r)$-form, $0\les r\les n$. Then, $\gf$ induces
a linear functional $T_{\gf}$ on the space of smooth $r$-forms with
compact supports defined by
\begin{equation}
T_{\gf}(\go)=\int_{\man}\gf\wedge\go.\label{eq:curr_by_form}
\end{equation}
Note that the integration above is valid since $\gf\wedge\go$ is
a smooth $n$-form having a compact support, and that the operation
is linear in $\go$.

A de Rham $r$-current is thus defined as a continuous linear functional
on the space of infinitely smooth differential $r$-forms, where the
continuity is defined by the following condition. Let $(\go_{a})$,
$a=1,2,\dots$, be a sequence of $r$-forms that are compactly supported
in the domain of a chart in $\man$ such that their components under
the chart and the partial derivatives of all orders of these components
converge uniformly to zero. Then, the linear functional $T$ is continuous
if 
\begin{equation}
\lim_{a\to\infty}T(\go_{a})=0.
\end{equation}

As a simple example of a singular de Rham $r$-current, let $w$ be
an $r$-vector at a point $m\in\man$. Then, $w$ induces the $r$-current
$T_{w}$ by setting
\begin{equation}
T_{w}(\go)=\go(m)(w).
\end{equation}
Clearly, the current $T_{w}$ is a generalization of the Dirac delta.

Given a current $T$, let $V$ be the largest open subset of $\man$
such that $T(\go)=0$ for any test form $\go$ supported in $V$.
Then the support of $T$ is defined as the complement of $V$.

The boundary, $\bdry T$, of an $r$-current, $T$, is the $(r-1)$-current
defined by
\begin{equation}
\bdry T(\go)=T(\dee\go),
\end{equation}
for any smooth compactly supported $(r-1)$-form $\go$. For the current
$T_{\gf}$ defined above, the boundary is given by 
\begin{equation}
\begin{split}\bdry T_{\gf}(\go) & =T_{\gf}(\dee\go),\\
 & =\int_{\man}\gf\wedge\dee\go,\\
 & =(-1)^{n-r}\left[\int_{\man}\dee(\gf\wedge\go)-\int_{\man}\dee\gf\wedge\go\right],\\
 & =(-1)^{n-r}\left[\int_{\bdry\man}\gf\wedge\go-\int_{\man}\dee\gf\wedge\go\right],\\
 & =(-1)^{n-r-1}\int_{\man}\dee\gf\wedge\go.
\end{split}
\end{equation}
Hence, 
\begin{equation}
\bdry T_{\gf}=(-1)^{n-r-1}T_{\dee\gf}.
\end{equation}
In light of this simple result, one may view the boundary of a current
as a generalization of the exterior differential.

The observations made above suggest that the non-smooth generalization
of the balance differential equation (\ref{eq:Diff-Balance-smooth})
in spacetime, and the fields appearing in it, will be
\begin{equation}
\bdry T=\vstd\label{eq:Ballance current-1}
\end{equation}
where $S$ is a zero-current on $\sptm$ representing the non-smooth
source, and $T$ is a one-current, representing the non-smooth flux
field. The zero-form\textemdash a real valued function\textemdash on
which $\bdry T$ and $S$ act is interpreted as a potential function
for the extensive property. The result of the evaluation of these
currents on the potential function is interpreted as physical power.

\myline

\smallskip{}

Using de Rham currents, it is possible to present surface growth as
a particular case of singular volumetric growth. Let $\reg\subset\sptm$
be an $(n+1)$-dimensional submanifold with boundary and let $\fourv J$
be a smooth $n$-form on $\reg$. Define the one-current $T$ on $\sptm$
by
\begin{equation}
T(\psi)=\int_{\reg}\fourv J\wedge\psi,
\end{equation}
for every smooth one-form $\psi$ compactly supported in $\sptm$.
Note that while the current is defined on $\sptm$, the integration
is performed only over $\reg$.

Computing the source current, $S$, we have
\begin{equation}
\begin{split}S(\gf)=\bdry T(\gf) & =T(\dee\gf),\\
 & =\int_{\reg}\fourv J\wedge\dee\gf,\\
 & =(-1)^{n}\int_{\reg}\dee(\fourv J\wedge\gf)+(-1)^{n+1}\int_{\reg}\dee\fourv J\wedge\gf,
\end{split}
\end{equation}
so that using Stokes's theorem,
\begin{equation}
S(\gf)=(-1)^{n}\int_{\bdry\reg}\fourv J\wedge\gf+(-1)^{n+1}\int_{\reg}\dee\fourv J\wedge\gf.
\end{equation}
The first integral on the right represents a source term supported
on the boundary $\bdry\reg$\textemdash surface growth. The second
integral on the right, represents smooth volumetric growth the source
density of which is $\dee\fourv J$.

In case $\fourv J$ is a closed form, only the surface growth term
remains. For example, if $\fourv J$ is an exact form, there is an
$(n-1)$-from $\ga$ such that $\fourv J=\dee\ga$, and $\dee\fourv J=0$,
automatically.

As a further particular example, consider the case where $\sptm=\reals\times\spc$
is a trivial bundle and $\spc$ can be covered by one chart $(x^{i})$.
Assume that $\fourv{J=\dee}x$ so that $\dee\fourv J=0$. Thus, the
action of the source current is
\begin{equation}
S(\gf)=(-1)^{n}\int_{\bdry\reg}\gf\,\dee x.\label{eq:SurfaceGrowthCurrents}
\end{equation}
It is noted that 
\begin{equation}
\fourv J=\dx=\bdry_{t}\contr(\dee t\wedge\dx).
\end{equation}
Hence, the vector field $\bdry_{t}$, induced by the volume element
$\dee t\wedge\dx$ according to (\ref{eq:v_contr-theta}), generates
the worldlines representing the material points.

In \cite{goldshtein2023notes} we present a simple example of such
surface growth.

\section{\label{sec:Whitey}Fractals as Currents and Flat Chains}

Some fractals may be described using the framework of the theory of
de Rham currents. Specifically, the theory of flat chains (see Whitney
\cite{Whitney1957}), which can be formulated in terms of de Rham
currents, offers such a natural geometric framework, as we will demonstrate
below.

We start reviewing the basic definitions concerning simplices and
polyhedral chains. Consider oriented $r$-simplices $\simp=[p_{0},\lisub p,r]$,
$p_{i}\in\affsp$, in an $n$-dimensional Euclidean space $\eucl^{n}$
(see, e.g., \cite{Segev_Book_2023}). The order of the points $p_{i}$
determines the orientation of a simplex. The simplex determined by
the same points as the simplex $\simp$ but having the reverse orientation
is $-\simp$. The $r$-volume of the simplex\textemdash its mass in
the terminology of geometric measure theory\textemdash will be denoted
by $\abs{\simp}$. An $r$-polyhedral chain $\chain$ is a formal
linear combination of $r$ simplices, that is,
\begin{equation}
\chain=\sum_{l=1}^{q}a_{l}\simp_{l},
\end{equation}
for some finite integer $q$. It is assumed that in case some of the
simplices in the linear combination are situated in the same hyperplane,
then, their interiors are disjoint. This can always be achieved by
subdivision of the simplices.

One identifies two chains if they have a common subdivision. Thus,
the mass of a polyhedral chain is 
\begin{equation}
\abs{\chain}=\sum_{l=1}^{q}\abs{a_{l}}\abs{\simp_{l}}.
\end{equation}
If $\chain=\sum_{\il=1}^{q}a_{\il}\simp_{\il}$ and $\chain'=\sum_{l'=1}^{q'}b_{l'}\tau_{l'}$,
using appropriate subdivisions one may represent $\chain$ and $\chain'$
in the forms
\begin{equation}
\chain=\sum_{\il''=1}^{q''}a'_{\il''}\simp'_{\il''},\qquad\chain'=\sum_{\il''=1}^{q''}b'_{\il''}\simp'_{\il''},\label{eq:two-chains}
\end{equation}
for one collection of simplices $\lisub{\simp'},{q''}$. Then, $\ga\chain+\ga'\chain'$
may be represented in the form
\begin{equation}
\ga\chain+\ga'\chain'=\sum_{\il''=1}^{q''}(\ga a'_{\il''}+\ga'b'_{\il''})\simp'_{\il''}.\label{eq:mutual-subdiv}
\end{equation}

For example, the $i$-th face of the simplex $\simp=[p_{0},\lisub p,r]$
is the simplex 
\begin{equation}
\tau_{i}=\sgn i[p_{0},\dots,\wh p_{i},\dots,p_{r}].\label{eq:boundary_simplex}
\end{equation}
and the boundary, $\bdry\simp$, of $\simp$ is the polyhedral chain
defined by
\begin{equation}
\bdry\simp=\sum_{\il=0}^{r}\tau_{\il}.
\end{equation}

The boundary of a polyhedral chain $\chain=\sum_{\il=1}^{q}a_{\il}\simp_{\il}$
is defined as 
\begin{equation}
\bdry\chain=\sum_{\il=1}^{q}a_{\il}\bdry\simp_{\il},
\end{equation}
by which the boundary operator is linear.

With these definitions, the collection of polyhedral $r$-chains may
be given the structure of a vector space, which is evidently infinite-dimensional.
An $r$-cochain is a linear functional defined on the space of $r$-polyhedral
chains. In order to ensure that the action of a cochain on a chain
is bounded by the mass of the chain and that the action on the boundary
$\bdry\chain$ of an $(r+1)$-chain $\chain$ is bounded by the mass
of $\chain$\textemdash properties suggested by flux theory where
the total flux through a chain is represented by the action of a cochain\textemdash led
Whitney to define the flat norm of a polyhedral $r$-chain $\chain$
by 
\begin{equation}
\norm{\chain}^{\flat}=\inf\{|\chain-\partial\D|+|\D|\},
\end{equation}
using all polyhedral $(r+1)$-chains, $\D$.

In particular, it follows immediately from the definition, that 
\begin{equation}
\begin{split}\norm{\bdry A}^{\flat} & \les\abs{\bdry A-\bdry A}+\abs A\\
 & \les\abs A,
\end{split}
\end{equation}
so that the flat norm of the boundary of a chain $A$ is always smaller
or equal to the mass of $A$.

The Banach space of flat $r$-chains is the completion of the space
of polyhedral $r$-chains relative to the flat norm. Thus, every flat
chain is the formal limit of the Cauchy sequence of polyhedral chains
relative to the flat norm.
\begin{rem}
The flux form in space is a co-chain, a continuous linear functional,
acting on the body chain (see \cite{Rodnay2003,Segev_Book_2023}).
Let $f_{i}:\reals\to\reals,i=0,1,...r$ be Lipschitz functions. By
the Rademaher theorem, Lipschitz functions are differentiable almost
everywhere at any simplex. A differential $r$-form, $\omega$, is
a Lipschitz form if it can be represented as 
\begin{equation}
\go=f_{0}\dee f_{1}\wedge\dee f_{2}\wedge\cdots.\wedge\dee f_{r}.
\end{equation}
This representation is not unique. Any Lipcshitz $r$-form is integrable
on any $r$-symplex and therefore on any flat chain. See \cite{Goldshtein82Forms,Goldshtein-Panenko2023}
for further details.
\end{rem}

\begin{example}
\label{exa:1-1}Let $A=\bdry C$, where $C$ is an equilateral triangle
of side $d$ in the Euclidean plane as illustrated in Figure \ref{fig:1-1}.
\begin{figure}
\begin{centering}
\includegraphics{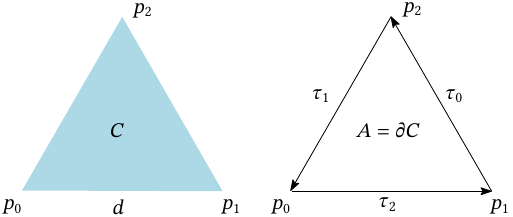}
\par\end{centering}
\caption{\label{fig:1-1}The triangle and its boundary as polyhedral chains
for Example \ref{exa:1-1}.}
\end{figure}
Then, 
\begin{equation}
\norm A^{\flat}=\norm{\bdry C}^{\flat}\les\abs C=\frac{\sqrt{3}}{4}d^{2}.
\end{equation}
It is noted that $\norm A^{\flat}$ is bounded by $d^{2}$ while $\abs A$
behaves like $d$.
\end{example}

\begin{example}
\label{exa:2-1}The von Koch snowflake is a standard example of a
fractal constructed, for example, as follows. One starts with the
$1$-simplex $B_{0}$ of unit length and at each step one sets,
\begin{equation}
\B_{i}=B_{0}+\sum_{j=1}^{i}\A_{j},\label{eq:Define_snowflake}
\end{equation}
as illustrated in Figure \ref{fig:2-1}.
\begin{figure}
\begin{centering}
\includegraphics[scale=0.67]{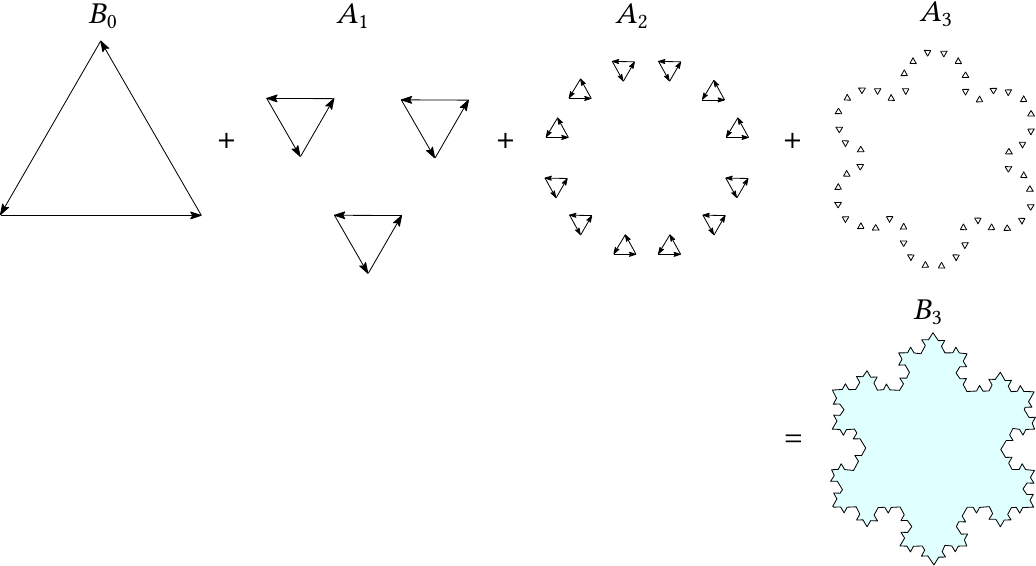}
\par\end{centering}
\caption{\label{fig:2-1}An illustration of the construction of the snowflake
as a flat chain for Example \ref{exa:2-1}.}
\end{figure}
 Now,
\begin{equation}
\A_{j}=\sum_{l=1}^{3\cdot4^{j-1}}\A_{jl},
\end{equation}
where $A_{jl}$ is the oriented boundary of a triangle of sides $3^{-i}$.
The triangles are situated in such a way, that at each step, the middle
third of the edge of the previous step is canceled.

The length of the resulting curve is unbounded and has the Hausdorff
dimension $\ln4/\ln3$. Since at each triangle adds two segments of
size $3^{-j}$, the total length added at the $j$-th step is 
\begin{equation}
3\cdot4^{j-1}\cdot2\cdot3^{-j}=\frac{3}{2}\left(\frac{4}{3}\right)^{j}.
\end{equation}

On the other hand, the flat norm of each $A_{jl}$ is bounded by the
area of the corresponding triangle, 
\begin{equation}
\fnorm{A_{jl}}\les\half\Bigl(\frac{\sqrt{3}}{2}3^{-j}\Bigr)3^{-j}=\frac{\sqrt{3}}{4}3^{-2j}.
\end{equation}
Thus,
\begin{equation}
\begin{split}\fnorm{A_{j}} & =\biggl\Vert\sum_{l=1}^{3\cdot4^{j-1}}\A_{jl}\biggr\Vert^{\flat},\\
 & \les\sum_{l=1}^{3\cdot4^{j-1}}\fnorm{\A_{jl}}\\
 & \les3\cdot4^{j-1}\frac{\sqrt{3}}{4}3^{-2j},\\
 & =\frac{3\sqrt{3}}{16}\left(\frac{4}{9}\right)^{j}.
\end{split}
\end{equation}
It follows that for $k>i$,
\begin{equation}
\begin{split}\fnorm{B_{k}-B_{i}} & =\biggl\Vert\sum_{j=i+1}^{k}\A_{j}\biggr\Vert^{\flat},\\
 & \les\sum_{j=i+1}^{k}\fnorm{\A_{j}},\\
 & \les\sum_{j=i+1}^{k}\frac{3\sqrt{3}}{16}\left(\frac{4}{9}\right)^{j}\\
 & \les\sum_{j=i+1}^{\infty}\frac{3\sqrt{3}}{16}\left(\frac{4}{9}\right)^{j},
\end{split}
\end{equation}
which is a convergent geometric series, the first term of which tends
to zero as $i\to\infty$. We conclude that this is a Cauchy sequence
and its limit, the snowflake, is a flat chain. Let us remark that
the resulting curve does not induce a flat cochain but a current of
different type. For example, integration on the resulting curve is
possible using the corresponding Hausdorff measure.
\end{example}

\begin{rem}
When constructed as above using equilateral triangles, the Hausdorff
dimension of the von Koch curve is $\ln4/\ln3>1$. Variants of the
standard von Koch snowflake are obtained when the typical triangles
added are general isosceles triangles (see an illustration in Figure
\ref{fig:alt-tria}). In such a case, the Hausdorff dimension of the
curve can be any number between $1$ and $2$.
\begin{figure}
\begin{centering}
\includegraphics[scale=0.75]{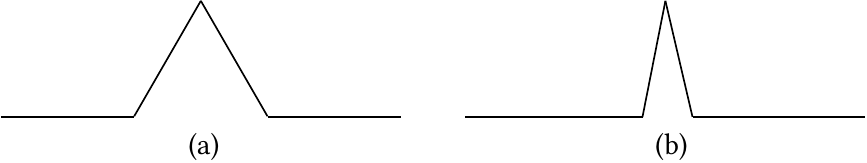}
\par\end{centering}
\caption{\label{fig:alt-tria}\protect \\
(a) A typical equilateral triangle in the construction of the snowflake.
\protect \\
(b) A different snowflake. The type of snowflake depends on the vertex
angle. Here (a) is a standard snowflake.}
\end{figure}
\end{rem}

To relate Whitney's theory of flat chains to de Rham currents we note
that Federer \cite{Federer1969}, defines flat chains as a particular
subspace of de Rham currents in open subsets of $\rn$. His construction
may be roughly described as follows. Given an open subset $U\subset\rn$,
let $K\subset U$ be a compact subset. The flat seminorm relative
to $K$ of an $r$-form $\gf$ on $U$ is defined as
\begin{equation}
\norm{\gf}_{K}^{\flat}:=\sup_{x\in K}\{\abs{\gf(x)},\abs{\dee\gf(x)}\},
\end{equation}
where one uses the natural metric structure of $\rn$ to evaluate
$\abs{\gf(x)}$ and $\abs{\dee\gf(x)}$. Then, a current supported
in $K$ is a flat chain if it is continuous relative to the norm $\norm{\cdot}_{K}^{\flat}$.

\section{\label{sec:Fractal-Growth}Fractal Growth}

In this section, we demonstrate growth processes that are associated
with fractal bodies. For example, let 
\begin{equation}
\reg=[0,1]\times\{0\}\subset\reals^{2},
\end{equation}
and let $Q\subset[0,1]$ be and arbitrary Cantor set. Then, the evolving
set
\begin{equation}
\reg_{t}=[0,1]\cup(Q\times[0,t])
\end{equation}
represents a growing fractal that may be applicable in the description
of percolation processes.

Next, we demonstrate a continuous growth process of a polyhedral chain
into a fractal. Specifically, we consider a continuous version of
the construction of the von Koch snowflake in Example \ref{exa:2-1}.
The notation and terminology of Example \ref{exa:2-1} will be used
below with the following modifications.

\begin{figure}
\begin{centering}
\includegraphics{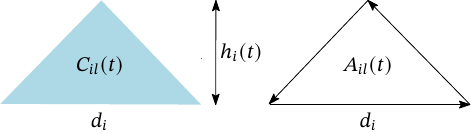}
\par\end{centering}
\caption{\label{fig:2-2}The triangles for the construction of the continuous
construction of the snowflake.}
\end{figure}

The triangles $A_{ij}$ are now time-dependent isosceles triangles.
The bases of the triangles are $d_{i}=3^{-i}$ and they are situated
in the middle thirds as before, but the heights of the triangles $h_{i}(t)$
are time-dependent (see Figure \ref{fig:2-2}).

For the time-interval $t\in[0,1]$ and $i=\oneto\infty$, we consider
the intervals $\gD_{i}=(t_{i-1},t_{i}]$ of lengths $\gD t_{i}=2^{-i}$
and the instances $t_{i}=t_{i-1}+\gD t_{i}$, where $t_{0}=0$. Thus,
\begin{equation}
t_{i}=\sum_{j=1}^{i}\gD t_{i}=1-2^{-i}.
\end{equation}
If $t\in\gD_{i}$, we set $\tau=t-t_{i-1}$.

We define the function $h:(0,1)\tto[0,1]$ such that
\begin{equation}
h(t)=2^{i}\tau.
\end{equation}
It follows that $h(t_{i})=2^{i}\gD_{i}=1$.

\begin{figure}
\begin{centering}
\includegraphics{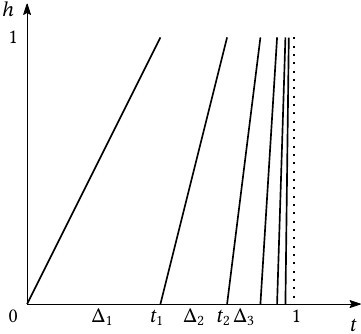}
\par\end{centering}
\caption{\label{fig:2-2-1}The function $h(t)$}
\end{figure}

Thus, we define the height of the triangle $h_{i}(t)$ as
\begin{equation}
h_{i}(t):=\frac{\sqrt{3}}{2}d_{i}h(t)=\frac{\sqrt{3}}{2}\left(\frac{2}{3}\right)^{i}\tau,\qquad t\in\gD_{i}
\end{equation}
so that
\begin{equation}
h_{i}(t)\overset{t\to t_{i-1}}{\tto}0,\qquad\text{and}\qquad h_{i}(t_{i})=\frac{\sqrt{3}}{2}d_{i}.
\end{equation}
We can formally extend $h_{i}$ by setting
\begin{equation}
h_{i}(t):=\begin{cases}
0, & \text{for }t\les t_{i-1},\\
1, & \text{for }t>t_{i}.
\end{cases}
\end{equation}

It is noted that the dependence of $i$ on $t$ can be obtained as
follows. If $t\in\gD_{i}$, then
\begin{equation}
i-1<-\log_{2}(1-t)\les i.
\end{equation}
Using the floor, $\left\lfloor \cdot\right\rfloor $, and ceiling
$\lceil\cdot\rceil$ functions, it is concluded that 
\begin{equation}
i=\lceil-\log_{2}(1-t)\rceil,\qquad i-1=\lfloor-\log_{2}(1-t)\rfloor.\label{eq:i(t)}
\end{equation}
Defining the function
\begin{equation}
I(t):=\lceil-\log_{2}(1-t)\rceil,
\end{equation}
Equation (\ref{eq:i(t)}) may be written as $i=I(t).$

Hence,
\begin{equation}
\begin{split}\tau & =t-t_{i-1},\\
 & =t-(1-2^{-(i-1)}),\\
 & =t-1+2^{-\lfloor-\log_{2}(1-t)\rfloor},
\end{split}
\end{equation}
and using $-\lfloor-x\rfloor=\lceil x\rceil$, 
\begin{equation}
\tau=t+2^{\lceil\log_{2}(1-t)\rceil}-1.
\end{equation}

At each interval $\gD_{i}$ the chains 
\begin{equation}
A_{i}(t)=\sum_{l=1}^{3\cdot4^{i-1}}A_{il}(t),
\end{equation}
with $A_{il}(t)$ as defined above in terms of $h_{i}(t)$, are added
continuously starting from zero height to equilateral triangles. Thus,
equation (\ref{eq:Define_snowflake}) is replaced by
\begin{equation}
\B(t)=B_{0}+\sum_{i=I(T)=1}^{\infty}\A_{i}(t)=B_{0}+\sum_{j=1}^{I(t)-1}\A_{j}+A_{I(t)}(t),
\end{equation}
where it is noted that by definition, only one term in the infinite
sum is different than zero, and we use the notation $A_{i}:=\A_{i}(t_{i})$.

As $t\to1$, the flat chain converges to the snowflake.

Finally, we note that integration theory of forms over chains on manifolds
and the corresponding Stokes theorem, implies that the setting of
(\ref{eq:SurfaceGrowthCurrents}) for surface growth applies here
also for any time $t_{0}<1$. We consider the chain $\reg$ defined
as follows. Let $C(t)$ be the chain bounded by $B(t)$, so that $B(t)=\bdry C(t),$
and set
\begin{equation}
\reg=\bigcup_{t_{0}>t'>t}\{(t',x)\mid x\in C(t),\;0\les t\les t_{0}\}.
\end{equation}
(Note that we have an artificial component of the boundary at $t_{0}$,
where all points are lost.)

This example demonstrates a continuous process of growth in which
the perimeter of the body grows without bound while the area of the
body remains bounded. Such a process may be advantageous when the
objective is increasing the transport through the boundary.

This heuristic description of ``added continuously starting from zero
height to equilateral triangles'' has a more precise formal interpretation.
For any $t$ and any $h_{i}(t)$, the new body $A_{i}(t)$ has a Lipschitz
(even piecewise linear) boundary, i.e., any of these bodies is a Lipschitz
domain. The procedure above is a mathematical description of an evolution
of Lipschitz domains into the snowflake (Figure \ref{fig:2-1}).

\label{sec:Evolution-to-Fractals}Evolution of Smooth Bodies to Fractals

\begin{figure}
\begin{centering}
\includegraphics[scale=0.8]{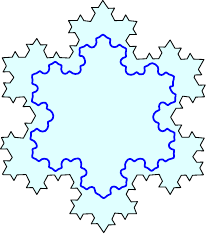}
\par\end{centering}
\caption{\label{fig:3-1}Illustrating the evolution of a smooth boundary into
a fractal one. (The black line represents the boundary of the snowflake,
and the blue line represents the smooth boundary of an approximation
in the growth process.)}
\end{figure}

In this section we present a construction that enables a mathematical
description of the evolution of two-dimensional bodies having smooth
boundaries into bodies with fractal boundaries (see an illustration
in Figure \ref{fig:3-1}).

Let $B(0,t)\subset\reals^{2}$ denote the open disk centered at the
origin of radius $t$, and let $G$ be an arbitrary simply connected
bounded open domain in $\reals^{2}$ representing a body, the boundary
of which may be fractal. By the Riemann mapping theorem, there is
a conformal diffeomorphism
\begin{equation}
\vph:B(0,2)\tto G.
\end{equation}
Consider the set 
\begin{equation}
\reg_{1}:=\vph\{\ol{B(0,1)}\},
\end{equation}
where the overline indicates closure. It follows that $\bdry\reg_{1}$
is smooth.

The domain $\reg_{1}$ represents the initial state of the growing
body, at time $t=1$ and we want to describe its evolution to a body
$\reg_{2}$ having a fractal boundary at time $t=2$. Thus, for $t\in[1,2)$,
we set
\begin{equation}
\reg_{t}:=\vph\{\ol{B(0,t)}\}.
\end{equation}
Let 
\begin{equation}
s_{t}:\reals^{2}\tto\reals^{2},\qquad s_{t}(x)=tx
\end{equation}
and consider the conformal diffeomorphism
\[
\vph_{t}:\reg_{1}\tto\reg_{t},\qquad\vph_{t}(x)=\vph\comp s_{t}\comp\vph^{-1}(x).
\]

By the theory of prime ends (see, for example, \cite{Prime_ends-Epstein},
\cite[Chap. 4]{Bracci2020}), the mapping $\vph$ can be extended
to the ``ideal'' Carathéodory boundary of $G$.\textsf{\textbf{\textcolor{brown}{}}}
If the \textsf{\textbf{\textcolor{brown}{}}}topological boundary
of $G$ is locally connected at any boundary point, the Carathéodory
boundary coincides with the topological boundary.

Regions in the plane with fractal boundaries, similar to the von Koch
snowflake above, satisfy this condition. In such a case, the fractal
boundary may be denoted as $\bdry\reg_{2}$. Thus, during the time
interval $[1,2]$, the smooth boundary $\bdry\reg_{1}$ evolved into
the fractal $\bdry\reg_{2}$.

\newcommand{\etalchar}[1]{$^{#1}$}

\end{document}